\documentclass[conference]{IEEEtran}

\usepackage{cite}

\ifCLASSINFOpdf

\else

\fi
\usepackage{amsmath}
\usepackage{cite}
\usepackage{graphicx}
\usepackage{caption}
\usepackage{subcaption}
\usepackage{float}
\usepackage{setspace}
\usepackage[font={small}]{caption}
\usepackage{multirow}
\usepackage{lipsum}
\input{epsf}
\begin{document}

\title{Synchrony in Neuronal Communication:\\An Energy Efficient Scheme}

\author{\IEEEauthorblockN{Siavash Ghavami$^{*}$, Vahid Rahmati$^{**}$, Farshad Lahouti${^*}$, Lars Schwabe${^{**}}$}
\IEEEauthorblockA{$^*$School of Electrical and Computer Engineering, University of Tehran, Tehran, Iran\\
$^{**}$ Faculty of Computer Science, University of Rostock, Rostock, Germany\\
Emails: s.ghavami@ut.ac.ir, vahid.rahmati@uni-rostock.de, lahouti@ut.ac.ir, lars.schwabe@uni-rostock.de}}

\maketitle

\begin{abstract}
We are interested in understanding the neural correlates of attentional processes using first principles.
Here we apply a recently developed first principles approach that uses transmitted information in bits per joule to quantify the energy efficiency of information transmission for an inter-spike-interval (ISI) code that can be modulated by means of the synchrony in the presynaptic population.
We simulate a single compartment conductance-based model neuron driven by excitatory and inhibitory spikes from a presynaptic population, where the rate and synchrony in the presynaptic excitatory population may vary independently from the average rate.
We find that for a fixed input rate, the ISI distribution of the post synaptic neuron depends on the level of synchrony and is well-described by a Gamma distribution for synchrony levels less than 50\%.
For levels of synchrony between 15\% and 50\% (restricted for technical reasons), we compute the optimum input distribution that maximizes the mutual information per unit energy.
This optimum distribution shows that an increased level of synchrony, as it has been reported experimentally in attention-demanding conditions, reduces the mode of the input distribution and the excitability threshold of post synaptic neuron.
This facilitates a more energy efficient neuronal communication.
\end{abstract}


\IEEEpeerreviewmaketitle

\section{Introduction}
%\lipsum[1-2]
Selective attention is affecting early stages of sensory processing~\cite{1} but the details of the underlying neuronal mechanisms are not fully uncovered yet. One theory proposes that the neural activity that represents the stimuli or events to be attended is selected through modification of its synchrony~\cite{2}. Detailed network modeling studies~\cite{3,4} built upon the idea that synchronous firing of neurons greatly affects the propagation of activity in network models~\cite{5,6} and could dynamically modulate the signal flow~\cite{7}.

The framework of information theory~\cite{8} has been successfully applied to early sensory coding in theoretical and modeling studies~\cite{9,10}, and the mutual information has been used as a measure to determine the information content of experimentally recorded responses in sensory systems~\cite{11}.
We are interested in understanding the role of synchrony in sensory information processing using a normative modeling approach.

More specifically, we adopt the notion that synchrony may have a modulatory role in neuronal signal processing~\cite{7} and consider the synchrony within a presynaptic population of neurons as an independent control parameter that adjusts the “channel characteristics” of the postsynaptic neuron. In other words, we conceptualize the postsynaptic neurons as a dynamically configurable communication channel through which information is communicated via an inter-spike-interval (ISI) code. We adopt the Berger-Levy theory of neural communication, which was recently proposed~\cite{12} and goes beyond information maximization approaches by postulating the maximization of capacity per unit cost (measured in bits per joule, bpj) as the biologically relevant objective for neurons~\cite{12,13}. In that line, the energy-efficiency has been suggested for retina~\cite{14} and cortex~\cite{15}, but normative modeling studies within the Berger-Levy theory remain rare~\cite{13,15,16,17}. In this paper, we ask the question "What is the best input distribution (over inter-spike-intervalls), which the maximize mutual information per unit cost in the said population of neurons and how this distribution is related to the level of synchrony?"

The role of synchrony in attention has been studied experimentally, e. g. in~\cite{7}, but here we apply mathematical modeling and simulation.
More specifically, we model a postsynaptic neuron based on the Hodgkin-Huxley model~\cite{23}.
Then, we use an information theoretic cost function to derive the optimal input distribution.
We vary independently the rate and synchrony in the presynaptic excitatory population of the conductance-based model neuron and characterize its input-output relation using simulations.
We consider the rates of excitatory neurons as representing the input and the ISI of the postsynaptic neuron as the output.
The probability of the single neuron's output (an ISI), conditioned on the input (the rate within the population), is determined experimentally as a function of the synchrony in the presynaptic excitatory population and fitted with parametric distributions.

We find that this probability distribution is well-described by a Gamma distribution for synchrony levels less than 50\%, which is normally reported in experimental measurements~\cite{20}.
For levels of synchrony between 15\% and 50\% (restricted for technical reasons) we compute the optimum input distribution that maximizes the mutual information per unit cost, which sheds light on how synchrony could affect the neuronal communication energy expenditure.

The remainder of this paper is organized as follows.
In Sections II and III, the modeling of neuronal synchrony and the neuronal communication channel are described.
In Section IV, we find the optimized input distribution for energy efficient communications.
Finally, in Section V, we conclude the paper.

\section{Modeling Neuronal Synchronization}
Our model is based on a single excitatory neuron which is driven by a homogeneous population of excitatory and inhibitory neurons. We modeled the postsynaptic neuron as a Hodgkin-Huxley-type (HH) model~\cite{23} with membrane potential $V$. Unlike the integrate-and-fire models, this biophysical model can generate spikes intrinsically by the following equation
\begin{equation}\label{eq:30}
{C_m}\frac{d}{{dt}}V\left( t \right) =  - {g_L}\left( {V\left( t \right) - {E_L}} \right) - \sum\limits_{{\mathop{int}} } {{J_{{\mathop{int}} }}\left( t \right)}  + {J_{net}}(t),
\end{equation}
where ${J_{{\mathop{int}} }}(t)$ denotes the active ionic current with Hodgkin-Huxley type kinetics, ${J_{net}}(t)$ is the synaptic current of the postsynaptic neuron, ${g_L}$ and ${E_L}$ are the leak conductance (${g_L} = 0.05~\frac{\rm {mS}}{{\rm c{m^2}}}$) and the reversal potential of the leak current (${E_L} =  - 65~{\rm{mV}}$), ${C_m}$ is the membrane capacitance (1~${\rm {\mu F} \mathord{\left/ {\vphantom {{\mu F} {\rm c{m^2}}}} \right. \kern-\nulldelimiterspace} {c{m^2}}}$), and $t$ is time. Each presynaptic neuron fires an independent Poisson spike train. We do not model the membrane potential of the presynaptic neurons, and consider their binary spiking activities; these spikes activate the presynaptic conductances, hence the synaptic input currents to the postsynaptic neuron are produced. The spike trains of the presynaptic neurons belonging to a subpopulation (excitatory/inhibitory) are generated with the same firing rate. To induce a controlled level of synchronicity between the presynaptic neurons, we model the occurrence of the synchronous events as another Poisson process. That generates an additional spike train with the rate determined by the synchronization rate between the presynaptic neurons. Here, we consider only synchronization in the excitatory subpopulation. We control the synchronicity in each subpopulation independent from its mean spiking activity. Therefore, in order to keep the mean activity constant between, e. g., the spiking activities in i) the absence ('old') and ii) presence ('new') of the synchronous events, the firing rate of each presynaptic neuron needs to be lowered in the case of synchronous spikes added to all presynaptic excitatory neurons: $\lambda_{ex}^{new} =\lambda_{ex}^{old}-S\lambda_{ex}^{syn}$, where $\lambda_{ex}^{old}$ is the firing rate in the absence of synchronous events, $\lambda_{ex}^{syn}$ is the rate of synchronous events, and $0<S<1$ denotes the fraction of the presynaptic neurons that are randomly chosen to participate in the synchronous events. This redefinition of the firing rate is applied to all neurons of excitatory subpopulation. Finally, the synchronized spike train can now be 'inserted' to the new (lower frequency) spike trains. In brief, an increase in the synchronization level can, in principle, yield larger fluctuations in the synaptic input currents, and thus in the postsynaptic membrane potential.

Balanced regimes are thought to play a crucial role in the transmission of information in cortical neurons in vivo.
For instance, recently it has been reported that these regimes can potentially promote both coding efficiency and energy efficiency~\cite{24}.
Accordingly, we also model a balanced activity regime of the excitatory and inhibitory neurons.
We parameterize the model in the following way to approximate such a balanced regime:
First, we define a constant input current $J^{ss}$, which in the absence of active ionic currents (see \eqref{eq:30}) leads to an asymptotic voltage of the model neuron's RC circuit close to the firing threshold of the full HH model neuron.
Then, we set this current equal to the summation of the means of all presynaptic excitatory and inhibitory currents ($J^{ss}_{ex}$ and $J^{ss}_{in}$), i. e. $J^{ss} = J^{ss}_{ex} + J^{ss}_{in}$.
We then find the desired parameter values of the corresponding synaptic input currents.
This results in constant synaptic conductance values ('weights') per synapse, that are independent of the synchronization level and the firing rate of individual presynaptic neurons.
Within our derivation, we make two biologically plausible assumptions: (i) the total firing rate of all presynaptic excitatory neurons is equal to that of inhibitory neurons, and (ii)  $J^{ss}_{ex} = 2J^{ss}$, i. e. without inhibition the excitatory drive would push the membrane potential way above the firing threshold.
Then, we fix the firing rate of the presynaptic inhibitory neurons (to 125 sp/s) and simulate the full HH-model for different rates of the presynaptic excitatory neurons, as well as different synchronization levels.
No additional background inputs or sources of noise were modeled or simulated.

We consider the level of the synchronization in the cell population as a controlling parameter for the neuronal communication channel.
In this line, an optimization problem is defined to find the optimum input distribution of the postsynaptic neuron to maximize the mutual information per unit cost for the neuronal communication channel.

\section{Modeling Neuronal Communication Channel}
\begin{figure*}
        \centering
        \begin{subfigure}[b]{0.85\textwidth}
        \vspace{-1.5\baselineskip}
                \includegraphics[width=\textwidth]{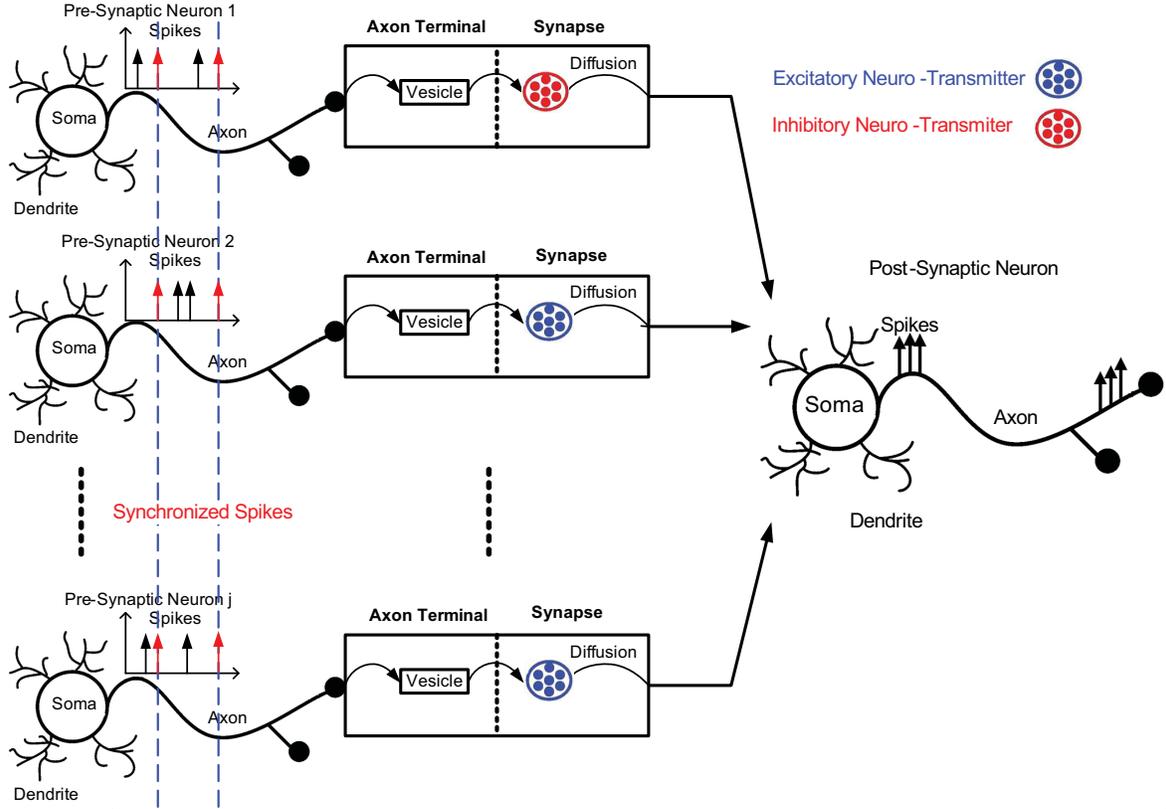}

        \end{subfigure}
        \vspace{-0.5\baselineskip}
        ~ %add desired spacing between images, e. g. ~, \quad, \qquad, \hfill etc.
        \caption{Illustration of the communication channel model. The excitatory and inhibitory neurons in the presynaptic population are firing spikes with rates ${\lambda _{ex}}$ and ${\lambda _{in}}$, respectively. These rates are encoded into ISIs sent through the channel (the set of synapses onto the postsynaptic neuron). Some spikes of the excitatory neurons are synchronized (blue arrows). Like any other spike of excitatory neurons, these synchronized spikes define ISIs, which encode ${\lambda _{ex}}$. For different levels of synchronization, and potentially different levels of inhibition, the channel itself changes its characteristics as reflected by different conditional distributions, i.e., $f\left( {\left. t \right|{\lambda _{ex}},\lambda _{in}^{\left( 1 \right)},{s^{\left( 1 \right)}}} \right) \ne f\left( {\left. t \right|{\lambda _{ex}},\lambda _{in}^{\left( 2 \right)},{s^{\left( 2 \right)}}} \right)$ for the same ${\lambda _{ex}}$  but different inhibitory rates $\lambda _{in}^{(1)}$  and $\lambda _{in}^{(2)}$ and/or synchronicities ${s^{(1)}}$ and ${s^{(2)}}$. Within this setting, the ${\lambda _{ex}}$ is communicated through the channel while ${\lambda _{in}}$ and $s$ control the channel characteristics.}
        \label{fig:1}
        \vspace{-1.5\baselineskip}
\end{figure*}

We consider the postsynaptic neuron as a communication channel.
The input of the communication channel are excitatory and inhibitory postsynaptic potential (EPSP and IPSP) intensities of neurons denoted by ${\lambda _{ex}}$ and ${\lambda _{in}}$. The output of the channel is the inter-spike interval (ISI) of postsynaptic neuron. The conditional probability of the output for given values of ${\lambda _{ex}}$ and ${\lambda _{in}}$ is controlled by level of synchrony within the excitatory population (See Fig.~\ref{fig:1}).
We model the conditional probability of ISIs, $f\left( {\left. t \right|{\lambda _{ex}},{\lambda _{in}},s} \right)$, using simulations of the Hodgkin-Huxley model.
The channel is assumed memory less and time invariant, i. e.
\begin{equation}
\begin{array}{l}
{f_{\left. {{T_1},...,{T_n}} \right|{\Lambda _{ex,1}},...,{\Lambda _{ex,n}},{\Lambda _{in,1}}...,{\Lambda _{in,n}},S}}\left( {\left. {{t_1},...,{t_n}} \right|} \right.\\
\left. {{\lambda _{ex,1}},...,{\lambda _{ex,n}},{\lambda _{in,1}},...,{\lambda _{in,n}},s} \right) = \\
\prod\limits_{k = 1}^n {{f_{\left. T \right|{\Lambda _{ex}},{\Lambda _{in}}}}\left( {\left. {{t_i}} \right|{\lambda _{ex,k}},...,{\lambda _{in,k}}} \right)} .
\end{array}
\end{equation}
The synchrony level is considered as a parameter of the channel.
We fix ${\lambda _{in}} = 125$ Hz and only vary ${\lambda _{ex}}$.
For brevity we drop ${\lambda _{in}}$ in $f\left( {\left. t \right|{\lambda _{ex}},{\lambda _{in}},s} \right)$ and denote it by $f\left( {\left. t \right|{\lambda _{ex}},s} \right)$.
%\lipsum[1-2]
\begin{figure*}[t!]
        \centering
        \begin{subfigure}[b]{0.49\textwidth}
        \vspace{-1\baselineskip}
                \caption{}
        \vspace{-0.5\baselineskip}
                \includegraphics[width=\textwidth]{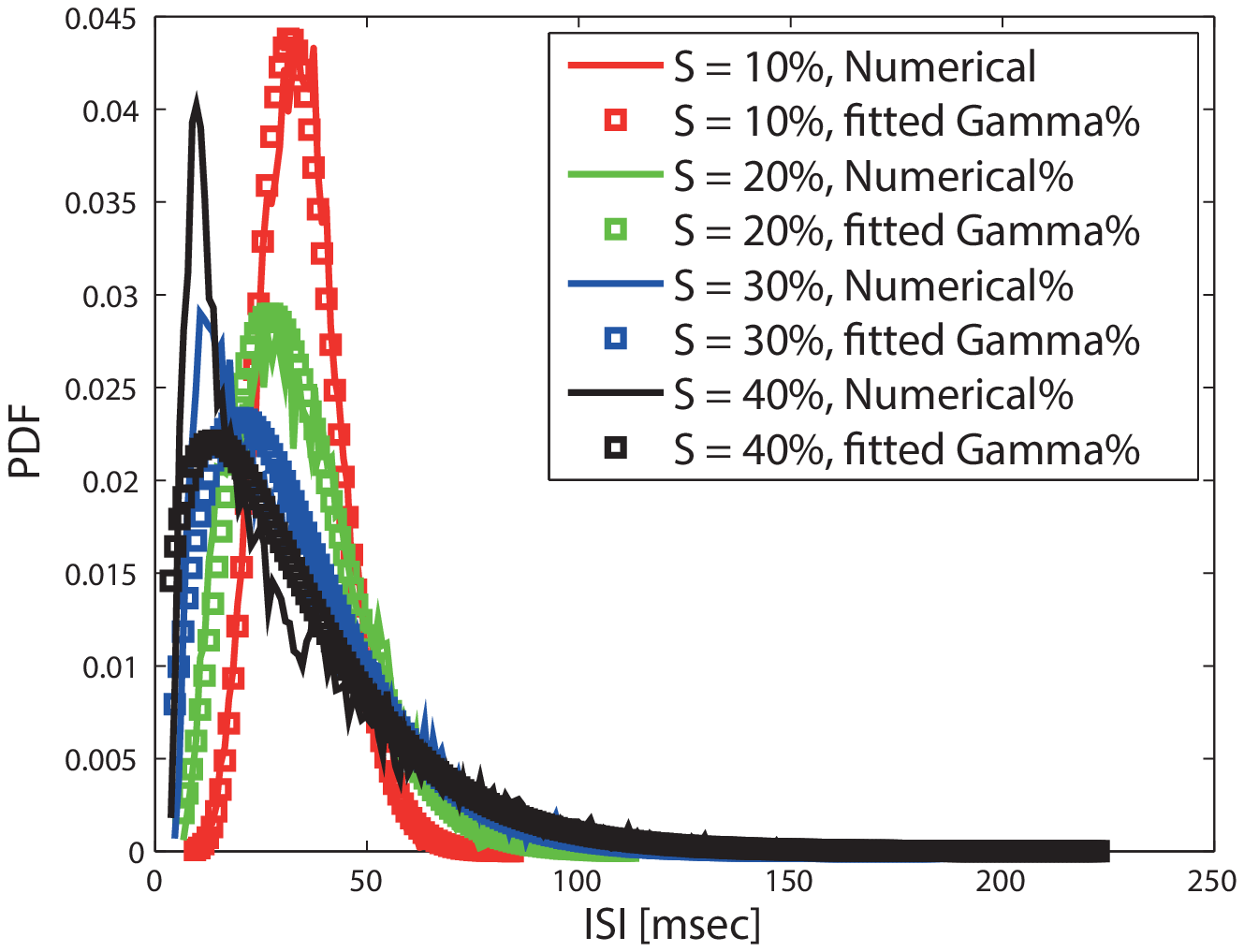}
                \label{fig:2a}
        \end{subfigure}%
        ~ %add desired spacing between images, e. g. ~, \quad, \qquad, \hfill etc.
%        \caption{}
        \begin{subfigure}[b]{0.49\textwidth}
        \vspace{-1\baselineskip}
                \caption{}
        \vspace{-0.5\baselineskip}
                \includegraphics[width=\textwidth]{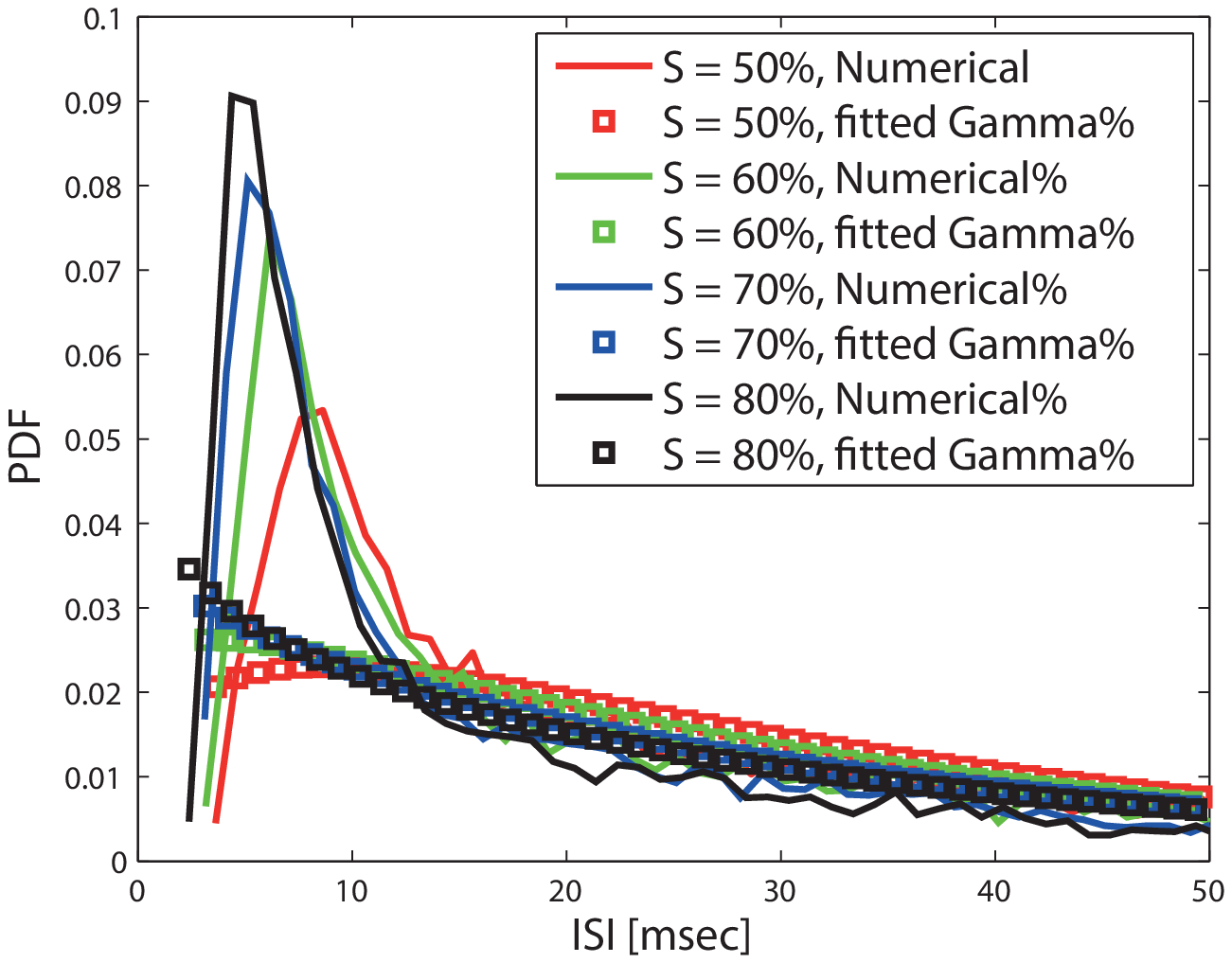}
                \label{fig:2b}
        \end{subfigure}
        \\
        \vspace{-1.5\baselineskip}
        \caption{Normalized histogram of ISI duration of simulated data and fitted Gamma distribution for different level of synchronization and ${\lambda _{ex}} = 36.0991$ Hz.}
        \label{fig:2}
        \vspace{-2\baselineskip}
\end{figure*}
\begin{figure*}[t!]
        \centering
        \begin{subfigure}[b]{0.49\textwidth}
        \vspace{-2\baselineskip}
                \caption{}
        \vspace{-0.5\baselineskip}
                \includegraphics[width=\textwidth]{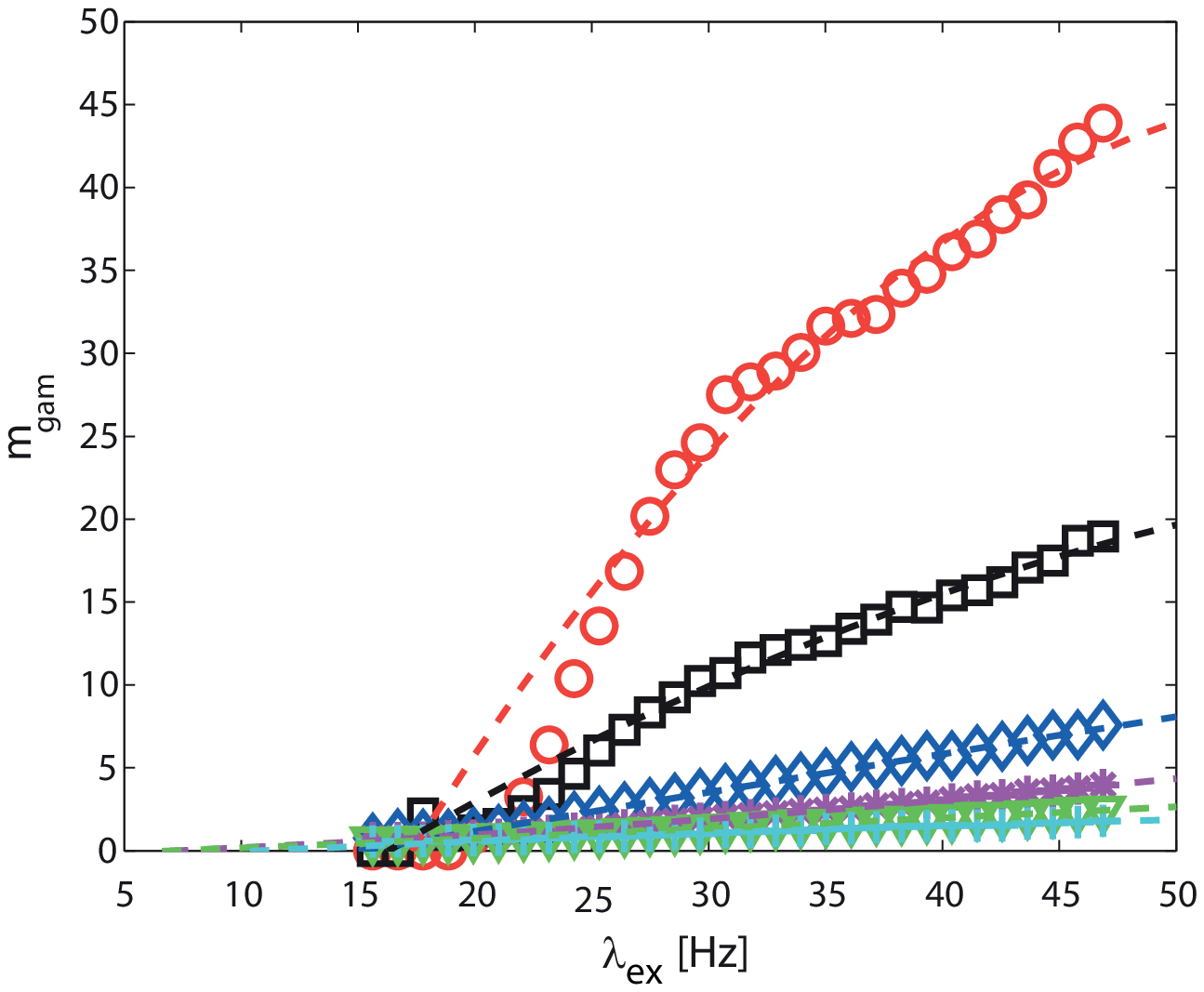}
                \label{fig:3a}
        \end{subfigure}%
        ~ %add desired spacing between images, e. g. ~, \quad, \qquad, \hfill etc.
        \begin{subfigure}[b]{0.49\textwidth}
                \caption{}
         \vspace{-0.5\baselineskip}
                \includegraphics[width=\textwidth]{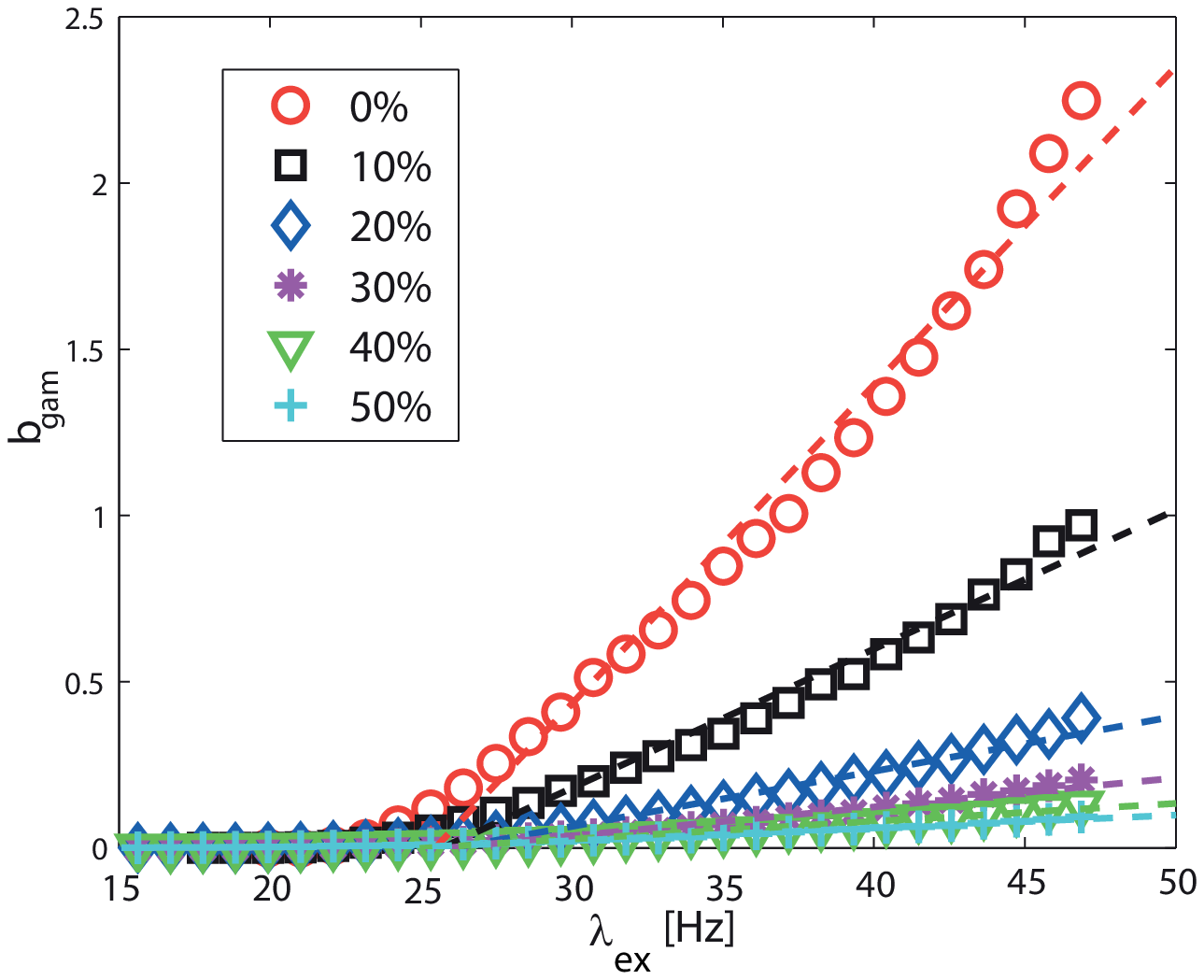}
                \label{fig:3b}
        \end{subfigure}
        \\
        \vspace{-1.5\baselineskip}
        \caption{Fits to the dependencies of the Gamma function parameters on the input rate. A. Quadratic fits (dash-line) to ${m_{gam}}$ (Markers)(as obtained from maximum likelihood fits to the simulated data) for  $s = 0\% $...50\%. B. Linear fit (dash-line) to ${b_{gam}}$ (Markers) for $s = 0\% $...50\%.}
        \label{fig:3}
        \vspace{-1.5\baselineskip}
\end{figure*}

The desired conditional probability is estimated using our simulation results.
Fig.\ref{fig:2a} and Fig.\ref{fig:2b} show the normalized histogram of ISI for different values of synchrony level.
As depicted in Fig.\ref{fig:2}, the Gamma distribution fits well to the obtained conditional ISI histograms, for $s$ less than 50\% and satisfies the Kolmogorov-Smirnov test with 5\% significance level.
Hence, we have
\begin{equation}\label{eq:4}
{f_{\left. T \right|{\Lambda _{ex}},S}}(\left. t \right|{\lambda _{ex}},s) = \frac{{{{\left( {{b_{gam}}} \right)}^{{m_{gam}}}}{t^{{m_{gam}} - 1}}{e^{ - {b_{gam}}t}}}}{{\Gamma \left( {{m_{gam}}} \right)}}u(t)
\end{equation}
where $b_{gam}$ and $m_{gam}$ are the scaling and shaping parameters of Gamma distribution which are obtained from maximum likelihood (ML) estimation.
We fit polynomial functions to the scaling and shaping parameters which are denoted by ${d^{(b)}}\left( {s,{\lambda _{ex}}} \right)$  (${d^{\left( m \right)}}\left( {s,{\lambda _{ex}}} \right)$) and are given by
\begin{equation}\label{eq:5}
d^{(b)}(s,{\lambda _{ex}}) = d_1^{(b)}(s){\lambda _{ex}} + d_0^{(b)}(s)
\end{equation}
\begin{equation}\label{eq:6}
d^{(m)}(s,{\lambda _{ex}}) = d_2^{(m)}(s)\lambda _{ex}^2 + d_1^{(m)}(s){\lambda _{ex}} + d_0^{(m)}(s)
\end{equation}
where $d_i^{\left( b \right)}\left( s \right)$, $i \in \left\{ {1,2} \right\}$ and $d_i^{\left( m \right)}\left( s \right)$, $i \in \left\{ {1,2,3} \right\}$ are coefficients of linear and quadratic functions.
The choice of these function types are due to our experiments for the best fit to the shaping and scaling parameters.
Fig.\ref{fig:3} shows the dependencies of the parameters of the Gamma distribution on the input rate.
Fig.\ref{fig:3a} and Fig.~\ref{fig:3b} show the shaping and scaling parameters of fitted gamma distribution to data, i.e. ${m_{gam}}$ and ${b_{gam}}$, and fitted polynomial functions to the shaping and scaling parameters, i.e. $d^{(m)}(s,{\lambda _{ex}})$ and $d^{(b)}(s,{\lambda _{ex}})$.
We found that ${m_{gam}}$ is well fit by a quadratic function.
To fit ${b_{gam}}$ a linear function is sufficient.

The synchrony level in real neurons is probably much less than 50\%~\cite{20}.
Our experiments reveal that the GEV distribution also fits well with the normalized histogram of ISI over all range of synchrony.
However, we opted for the Gamma distribution instead, since it allows for an efficient solution of problem \eqref{eq:2} as we shall see below.
\vspace{-0.5\baselineskip}

\section{Optimized Input Distribution for Energy Efficient Communications}
We seek the optimum distribution of ${\lambda _{ex}}$ for maximizing the average mutual information given a synchronicity level per unit cost in neurons.
The associated optimization problem is described as follows
\begin{equation}\label{eq:2}
\begin{array}{l}
{I_{bpj}} = {\rm{      }}\mathop {\max }\limits_{{F_{\left. {{\Lambda _{ex}}} \right|S}}\left( {\left. {{\lambda _{ex}}} \right|s} \right)} \frac{I}{{E\left( {e(t)} \right)}},\\
{\rm{          }}s.t.{\rm{  }}{F_{\left. {{\Lambda _{ex}}} \right|S}}\left( {\left. {{\lambda _{ex}}} \right|s} \right) = \Pr \left( {\left. {{\Lambda _{ex}} < {\lambda _{ex}}} \right|S = s} \right),
\end{array}
\end{equation}
also $e(t) = {C_0} + {C_1}E(T),$ is the energy expenditure function of neuron during the ISI of duration $T$, ${C_0}$ and ${C_1}$ are constants~\cite{12}.
Moreover, ${F_{\left. {{\Lambda _{ex}}} \right|S}}\left( {\left. {{\lambda _{ex}}} \right|s} \right)$ denotes the cumulative distribution function (CDF) of ${\lambda _{ex}}$ for given value of $s$. In~\eqref{eq:2}, $I$ is the average mutual information for given value of synchrony level, and is given by
\begin{equation}\label{eq:3}
I = \frac{1}{N}\mathop {\lim }\limits_{N \to \infty } I\left( {\left. {{\Lambda _{ex,1}},...,{\Lambda _{ex,N}};{T_1},...,{T_N}} \right|S} \right),
\end{equation}
where, ${\Lambda _{ex,i}},{T_i},i \in \left\{ {1,...,N} \right\}$ denote the EPSP intensity and ISI, respectively, and $N$ denotes the number of spikes of postsynaptic neuron during time $T$.
%\vspace{-0.5\baselineskip}
For solving the optimization problem~\eqref{eq:2}, we model a communication channel (inputs are firing rates, outputs are ISIs) by considering the synchrony level as control parameters of the channel.
A simpler development in the case of a leaky integrate and fire model neuron is available in~\cite{12} without considering synchrony and inhibitory firing rate.

In this Section, we determine an equivalent problem for solving the optimization problem in~\eqref{eq:2}, which is easier to solve than the original problem.
We can find a closed form expression for optimization problem in~\eqref{eq:2} with ${f_{\left. T \right|{\Lambda _{ex}},S}}(\left. t \right|{\lambda _{ex}},s)$  in~\eqref{eq:4} and a range of synchronicity between 15\% to 50\%.
Briefly, in this case, the equivalent problem reduces to finding the CDF of ISIs, denoted by ${F_{\left. T \right|S}}\left( {\left. t \right|s} \right)$, which maximizes $h\left( {\left. T \right|s} \right)$ or the ISI entropy, subject to the constraints.
Upon obtaining ${F_{\left. T \right|S}}\left( {\left. t \right|s} \right)$ for feasible values of the constraints in~\eqref{eq:2}, we then seek the corresponding optimized ${F_{\left. {{\Lambda _{ex}}} \right|S}}\left( {\left. {{\lambda _{ex}}} \right|s} \right)$.
In line with~\cite{12} using~\eqref{eq:4} in~\eqref{eq:2} and by exploiting Lagrange function and due to linearity of $d^{(b)}(s,{\lambda _{ex}})$ in terms of $\lambda _{ex}$ (more details are omitted for brevity), the optimization problem in \eqref{eq:2} can be simplified to the following optimization problem
\begin{subequations} \label{eq:7}
\begin{align}
\label{eq:7a}
&\mathop {\max }\limits_{{F_{\left. T \right|S}}\left( {\left. t \right|s} \right)} h\left( {\left. T \right|s} \right),\\ \nonumber
s.t.&\\
&{F_{\left. T \right|S}}\left( {\left. t \right|s} \right) = \Pr \left( {\left. {T < t} \right|s} \right),\\
&E\left( {\left. T \right|s} \right) = {g_0},\\
&E\left( {\left. {{{\log }_e}T} \right|s} \right) = {g_1},
\end{align}
\end{subequations}
where the constraint on $E\left( {\left. T \right|s} \right)$ comes from the expression for the energy, which is of the form $e\left( T \right) = {C_0} + {C_1}T$.
The constraint from $E\left( {\left. {{{\log }_e}T} \right|s} \right)$ comes from the expression for the mutual information between two successive ISI~\cite{12}.
The optimized distribution for ${f_{\left. T \right|S}}(\left. t \right|s)$ is a Gamma distribution~\cite{21} as
\begin{equation}\label{eq:8}
{f_{\left. T \right|S}}(\left. t \right|s) = \frac{{{\beta ^\kappa }{t^{\kappa  - 1}}{e^{ - \beta t}}}}{{\Gamma (\kappa )}}u(t).
\end{equation}
where $\beta$ and $\kappa$ are shaping and scaling parameters of the Gamma distribution of the ISIs obtained from the constraints ${g_1} = {\kappa  \mathord{\left/
 {\vphantom {\kappa  \beta }} \right.
 \kern-\nulldelimiterspace} \beta }$ and ${g_0} = \psi \left( \kappa  \right) - {\rm{log}}\left( \beta  \right)$,  where ${\rm{log}}\left(  \cdot  \right)$  and $\psi \left(  \cdot  \right)$ are the natural logarithm and digma functions~\cite{22}.

The marginal ISI distribution is obtained by marginalizing over the input rate,
\begin{equation}\label{eq:9}
{f_{\left. T \right|S}}(\left. t \right|s) = \int {d{\lambda _{ex}}{f_{\left. {{\Lambda _{ex}}} \right|S}}\left( {\left. {{\lambda _{ex}}} \right|s} \right)} {f_{\left. T \right|{\Lambda _{ex}},S}}\left( {\left. t \right|{\lambda _{ex}},s} \right).
\end{equation}

Based on this formula, we can compute ${f_{\left. {{\Lambda _{ex}}} \right|S}}(\left. {{\lambda _{ex}}} \right|s)$ from solving the following integral equation
\begin{equation}\label{eq:10}
\begin{array}{l}
{f_{\left. T \right|S}}\left( {\left. t \right|s} \right) = \int {{f_{\left. {{\Lambda _{ex}}} \right|S}}} \left( {\left. {{\lambda _{ex}}} \right|s} \right) \times \\
\frac{{{{\left( {{d^{(b)}}\left( {s,{\lambda _{ex}}} \right)} \right)}^{{d^{(m)}}\left( {s,{\lambda _{ex}}} \right)}}{t^{{d^{(m)}}\left( {s,{\lambda _{ex}}} \right) - 1}}{e^{ - {d^{(b)}}\left( {s,{\lambda _{ex}}} \right)t}}}}{{\Gamma \left( {{d^{\left( m \right)}}\left( {s,{\lambda _{ex}}} \right)} \right)}}u(t)d{\lambda _{ex}}\\
 = \frac{{{\beta ^\kappa }{t^{\kappa  - 1}}{e^{ - \kappa t}}}}{{\Gamma (\kappa )}}u(t),
\end{array}
\end{equation}
The optimum distribution of ${\lambda _{ex}}$ is obtained by following theorem.

\textbf{Theorem 1}. The optimum distribution of EPSP intensity for a given synchrony level, $s$, in the context of problem~\eqref{eq:2}, is given by
\begin{equation}\label{eq:11}
\begin{array}{l}
{f_{\left. {{\Lambda _{ex}}} \right|S}}\left( {\left. {{\lambda _{ex}}} \right|s} \right) = {\beta ^\kappa }d_1^{(b)}\left( s \right)\frac{{\Gamma \left( {d_0^{(m)}\left( s \right)} \right)}}{{\Gamma \left( \kappa  \right)\Gamma \left( {d_0^{(m)}\left( s \right) - \kappa } \right)}}\\
.\frac{{{{\left( {{\lambda _{ex}}d_1^{(b)}\left( s \right) - \beta  + d_0^{(b)}\left( s \right)} \right)}^{d_0^{(m)}\left( s \right) - \kappa  - 1}}}}{{{{\left( {{\lambda _{ex}}d_1^{(b)}\left( s \right) + d_0^{(b)}\left( s \right)} \right)}^{d_0^{(m)}\left( s \right)}}}}\\
.u\left( {{\lambda _{ex}}d_1^{(b)}\left( s \right) - \beta  + d_0^{(b)}\left( s \right)} \right)
\end{array}
\end{equation}
where $\Gamma \left( . \right)$ denotes the Gamma function.

\emph{Proof}. See Appendix A.

In Fig.~\ref{fig:4}, the results of the optimization problem is shown for $s = $
30\%, 40\% and 50\% with $E\left( {\left. t \right|s} \right) = 100 {\rm{ msec}}$  and $E{\rm{(}}\left. {{\rm{log(}}T{\rm{)}}} \right|s{\rm{)}} =  - 3.51$.
By increasing the synchrony level, the mode of EPSP intensity ${\lambda _{ex}}$, corresponding to the peak value of ${f_{\left. {{\Lambda _{ex}}} \right|S}}\left( {\left. {{\lambda _{ex}}} \right|s} \right)$, is reduced.
Moreover, higher synchronicity reduces the minimum value of ${\lambda _{ex}}$ with non-zero probability.
This shows that according to the optimized energy efficient strategy in~\eqref{eq:2}, enhanced synchrony reduces the excitation threshold of the post synaptic neuron.

\begin{figure}
        \centering
        \vspace{-2\baselineskip}
        \begin{subfigure}[b]{0.49\textwidth}
                \includegraphics[width=\textwidth]{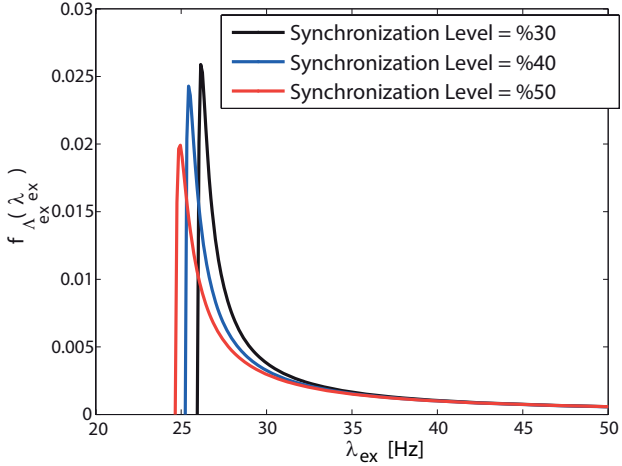}

        \end{subfigure}
        \vspace{-1.5\baselineskip}
        ~ %add desired spacing between images, e. g. ~, \quad, \qquad, \hfill etc.
        \caption{Optimum distribution of ${f_{\left. {{\Lambda _{ex}}} \right|S}}\left( {\left. {{\lambda _{ex}}} \right|s} \right)$  for optimization problem~\eqref{eq:2} with different values of synchrony level.}
        \label{fig:4}
        \vspace{-2.5\baselineskip}
\end{figure}
\vspace{-1\baselineskip}

\section{Concluding Remarks}
We investigated the role of neuronal synchrony from a communication-theoretic point of view by modeling a neuros as a communication channel with synchrony as the channel's control parameter.
The excitatory post synaptic potential (EPSP) intensity and the inter spike interval (ISI) are the input and the output of the channel model.
Our simulation results showed that the conditional probability of the neuronal communication channel is well fitted with the Gamma distribution for synchrony levels less than 50\%.
The optimum distribution of ${\lambda _{ex}}$  for a given value of $s$ is analytically obtained and shows that increasing the level of synchrony reduces the mode of EPSP intensity distribution and the threshold of excitation.

Synchrony of presynaptic neurons is observed during the attention process.
Our results now present another interpretation of this experimental observation: Instead of synchronicity being the carrier of information, it may primarily control the information flow in an energy efficient way.

\vspace{-2\baselineskip}
\section{Appendix A. Proof Of Theorem 1}
\begin{figure}
        \centering
        \vspace{-2\baselineskip}
        \begin{subfigure}[b]{0.49\textwidth}
                \includegraphics[width=\textwidth]{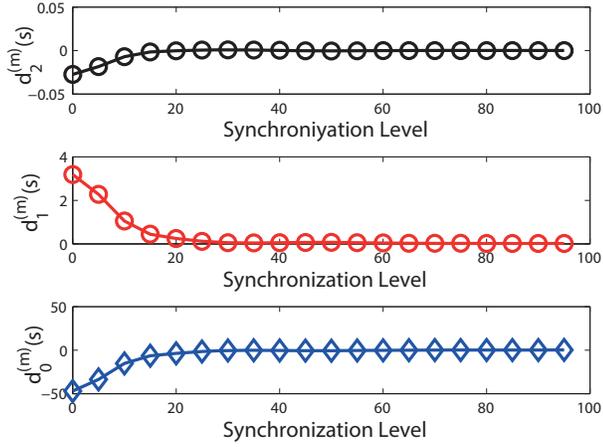}

        \end{subfigure}
        \vspace{-2\baselineskip}
        ~ %add desired spacing between images, e. g. ~, \quad, \qquad, \hfill etc.
        \caption{ $d^{(i)}_m(s)$, $i \in 0,1,2$ in terms of synchronization level.}
        \label{fig:5}
        \vspace{-2\baselineskip}
\end{figure}

By replacing ${d^{(b)}}\left( {s,{\lambda _{ex}}} \right)$  and  ${d^{\left( m \right)}}\left( {s,{\lambda _{ex}}} \right)$ from~\eqref{eq:5} and~\eqref{eq:6} in~\eqref{eq:9}, we have
\begin{equation}\label{eq:12}
\begin{array}{l}
\int\limits_0^\infty  {{f_{\left. {{\Lambda _{ex}}} \right|S}}} \left( {\left. {{\lambda _{ex}}} \right|s} \right){t^{d_2^{(m)}\left( s \right)\lambda _{ex}^2 + d_1^{(m)}\left( s \right){\lambda _{ex}} + d_0^{(m)}\left( s \right) - 1}} \times \\
{e^{ - \left( {d_1^{(b)}\left( s \right){\lambda _{ex}} + d_0^{(b)}\left( s \right)} \right)t}} \times \\
\frac{{{{\left( {d_1^{(b)}\left( s \right){\lambda _{ex}} + d_0^{(b)}\left( s \right)} \right)}^{d_2^{(m)}\left( s \right)\lambda _{ex}^2 + d_1^{(m)}\left( s \right){\lambda _{ex}} + d_0^{(m)}\left( s \right)}}}}{{\Gamma (d_2^{(m)}\left( s \right)\lambda _{ex}^2 + d_1^{(m)}\left( s \right){\lambda _{ex}} + d_0^{(m)}\left( s \right))}}d{\lambda _{ex}}\\
 = \frac{{{\beta ^\kappa }{t^{\kappa  - 1}}{e^{ - \beta t}}{e^{d_0^{(b)}\left( s \right)t}}}}{{\Gamma (\kappa )}}.
\end{array}
\end{equation}
By a change of variable $v = d_1^{(b)}\left( s \right){\lambda _{ex}}$, we have
\begin{equation}\label{eq:13}
\begin{array}{l}
\frac{1}{{d_1^{(b)}\left( s \right)}}\int {{f_{\left. {{\Lambda _{ex}}} \right|S}}} \left( {\left. {\frac{v}{{d_1^{(b)}\left( s \right)}}} \right|s} \right){e^{ - vt}} \times \\
\frac{{{{\left( {v + d_0^{(b)}\left( s \right)} \right)}^{d_2^{(m)}\left( s \right){{\left( {\frac{v}{{d_1^{(b)}\left( s \right)}}} \right)}^2} + d_1^{(m)}\left( s \right)\left( {\frac{v}{{d_1^{(b)}\left( s \right)}}} \right) + d_0^{(m)}\left( s \right)}}}}{{\Gamma \left( {d_2^{(m)}\left( s \right){{\left( {\frac{v}{{d_1^{(b)}\left( s \right)}}} \right)}^2} + d_1^{(m}\left( s \right)\left( {\frac{v}{{d_1^{(b)}\left( s \right)}}} \right) + d_0^{(m)}\left( s \right)} \right)}} \times \\
{t^{d_2^{(m)}\left( s \right){{\left( {\frac{v}{{d_1^{(b)}\left( s \right)}}} \right)}^2} + d_1^{(m)}\left( s \right)\left( {\frac{v}{{d_1^{(b)}\left( s \right)}}} \right) + d_0^{(m)}\left( s \right) - 1}}dv\\
 = \frac{{{\beta ^\kappa }{t^{\kappa  - 1}}{e^{ - \beta t}}{e^{d_0^{(b)}\left( s \right)t}}}}{{\Gamma (\kappa )}}u(t).
\end{array}
\end{equation}
By simplification, we have
\begin{equation}\label{eq:14}
\begin{array}{l}
\frac{1}{{d_1^{(b)}\left( s \right)}}\int {{f_{\left. {{\Lambda _{ex}}} \right|S}}} \left( {\left. {\frac{v}{{d_1^{(b)}\left( s \right)}}} \right|s} \right){e^{ - vt}} \times \\
\frac{{{{\left( {v + d_0^{(b)}\left( s \right)} \right)}^{{{d'}_2}\left( s \right){v^2} + {{d'}_1}\left( s \right)v + d_0^{(m)}\left( s \right)}}}}{{\Gamma \left( {{{d'}_2}\left( s \right){v^2} + {{d'}_1}\left( s \right)v + d_0^{(m)}\left( s \right)} \right)}}{t^{{{d'}_2}\left( s \right){v^2} + {{d'}_1}\left( s \right)v + d_0^{(m)}\left( s \right) - 1}}dv\\
 = \frac{{{\beta ^\kappa }{t^{\kappa  - 1}}{e^{ - \beta t}}{e^{d_0^{(b)}\left( s \right)t}}}}{{\Gamma (\kappa )}}u(t),
\end{array}
\end{equation}
where, ${d'_2}\left( s \right) = d_2^{(m)}\left( s \right)/{\left( {d_1^{(b)}\left( s \right)} \right)^2}$ , ${d'_1}\left( s \right) = {{d_1^{(m)}\left( s \right)} \mathord{\left/
 {\vphantom {{d_1^{(m)}\left( s \right)} {{{\left( {d_1^{(b)}\left( s \right)} \right)}^2}}}} \right.
 \kern-\nulldelimiterspace} {{{\left( {d_1^{(b)}\left( s \right)} \right)}^2}}}$. A closed form solution to the above integral equation is illusive. Our simulation results (See Fig.~\ref{fig:5}) shows that for $s > 15\% $, $d_2^{(m)}\left( s \right)$ and $d_1^{(m)}\left( s \right)$ are almost zero. Hence, we can write \eqref{eq:14} as
\begin{equation}\label{eq:15}
\begin{array}{l}
\frac{1}{{d_1^{(b)}\left( s \right)}}\int {{f_{\left. {{\Lambda _{ex}}} \right|S}}} \left( {\left. {\frac{v}{{d_1^{(b)}\left( s \right)}}} \right|s} \right){e^{ - vt}} \times \\
\frac{{{{\left( {v + d_0^{(b)}\left( s \right)} \right)}^{d_0^{(m)}\left( s \right)}}{t^{d_0^{(m)}\left( s \right) - 1}}}}{{\Gamma (d_0^{(m)}\left( s \right))}}dv = \frac{{{\beta ^\kappa }{t^{\kappa  - 1}}{e^{ - \beta t}}{e^{d_0^{(b)}\left( s \right)t}}}}{{\Gamma (\kappa )}}u(t).
\end{array}
\end{equation}
Noting definition of Laplace transform, we have
\begin{equation}\label{eq:16}
\begin{array}{l}
\frac{1}{{d_1^{(b)}\left( s \right)}}{{\cal L}}\left( {{f_{\left. {{\Lambda _{ex}}} \right|S}}\left( {\left. {\frac{v}{{d_1^{(b)}\left( s \right)}}} \right|s} \right)\frac{{{{\left( {v + d_0^{(b)}\left( s \right)} \right)}^{d_0^{(m)}\left( s \right)}}{t^{d_0^{(m)}\left( s \right) - 1}}}}{{\Gamma \left( {d_0^{(m)}\left( s \right)} \right)}}} \right)\\
 = \frac{{{\beta ^\kappa }{t^{\kappa  - 1}}{e^{ - \beta t}}{e^{d_0^{(b)}\left( s \right)t}}}}{{\Gamma (\kappa )}}u(t).
\end{array}
\end{equation}
Using inverse Laplace transform, we have
\begin{equation}\label{eq:17}
\begin{array}{l}
\frac{1}{{d_1^{(b)}\left( s \right)}}{f_{\left. {{\Lambda _{ex}}} \right|S}}\left( {\left. {\frac{v}{{d_1^{(b)}\left( s \right)}}} \right|s} \right)\frac{{{{\left( {v + d_0^{(b)}\left( s \right)} \right)}^{d_0^{(m)}\left( s \right)}}}}{{\Gamma \left( {d_0^{(m)}\left( s \right)} \right)}} = \\
{\beta ^\kappa }{\left( {v - \beta  + d_0^{(b)}\left( s \right)} \right)^{d_0^{(m)}\left( s \right) - \kappa  - 1}}u\left( {v - \beta  + d_0^{(b)}\left( s \right)} \right),
\end{array}
\end{equation}
hence, we obtain
\begin{equation}\label{eq:17}
\begin{array}{l}
{f_{\left. {{\Lambda _{ex}}} \right|S}}\left( {\left. {\frac{v}{{d_1^{(b)}\left( s \right)}}} \right|s} \right) = {\beta ^\kappa }d_1^{(b)}\left( s \right)\frac{{\Gamma (d_0^{(m)}\left( s \right))}}{{\Gamma \left( \kappa  \right)\Gamma \left( {d_0^{(m)}\left( s \right) - \kappa } \right)}}\\
\frac{{{{\left( {v - \beta  + d_0^{(b)}\left( s \right)} \right)}^{d_0^{(m)}\left( s \right) - \kappa  - 1}}}}{{{{\left( {v + d_0^{(b)}\left( s \right)} \right)}^{d_0^{(m)}\left( s \right)}}}}u\left( {v - \beta  + d_0^{(b)}\left( s \right)} \right).
\end{array}
\end{equation}
Replacing $v = d_1^{(b)}\left( s \right){\lambda _{ex}}$, we have
\begin{equation}\label{eq:17}
\begin{array}{l}
{f_{\left. {{\Lambda _{ex}}} \right|S}}\left( {\left. {{\lambda _{ex}}} \right|s} \right) = {\beta ^\kappa }d_1^{(b)}\left( s \right)\frac{{\Gamma \left( {d_0^{(m)}\left( s \right)} \right)}}{{\Gamma \left( \kappa  \right)\Gamma \left( {d_0^{(m)}\left( s \right) - \kappa } \right)}} \times \\
\frac{{{{\left( {{\lambda _{ex}}d_1^{(b)}\left( s \right) - \beta  + d_0^{(b)}\left( s \right)} \right)}^{d_0^{(m)}\left( s \right) - \kappa  - 1}}}}{{{{\left( {{\lambda _{ex}}d_1^{(b)}\left( s \right) + d_0^{(b)}\left( s \right)} \right)}^{d_0^{(m)}\left( s \right)}}}} \times \\
u({\lambda _{ex}}d_1^{(b)}\left( s \right) - \beta  + d_0^{(b)}\left( s \right))
\end{array}
\end{equation}

\bibliographystyle{IEEEtran}
\bibliography{IEEE_IWCIT}

\end{document}